# Wound pH depends on actual wound size

T. Sirkka*, J.B. Skiba** and S. P. Apell*

*   Department of Physics and Gothenburg Physic Centre,
    Chalmers University of Technology, SE-412 96 Göteborg, Sweden
** MedMinding LLC, Chandler, AZ 85226, USA.

**Abstract:**
Wound healing is an intricate process that involves many types of cells, reaction pathways as well as chemical, physical and electrical cues. Since biochemical reactions and physiological events are pH-dependent we study here pH as an important major characteristic of the wound healing process in the presence of endogenous and exogenous electric fields. Our model gives the spatial pH distribution in a wound. In particular we isolate a number of dimensionless quantities which sets the length, energy and time scales governing the wound healing process and which can be experimentally tested. Most interesting finding is that wound pH depends on actual wound size.

## pH in wounds

As a consequence of a tissue trauma resulting in a wound, a set of delicate processes are initiated to restore homeostasis in the damaged area. Wound healing is a field of study that must have a history as long as that of mankind. Some aspects of the healing process are rather well studied and understood by now. Due to the complexity of the process there are however many that remain yet uncharted or unappreciated. The shortcomings in our understanding are reflected in wound care where impaired wound healing and chronic wounds still are a major problem that in total have immense social and economic consequences and ultimately challenges us to design new sensing, imaging and treatment techniques [1,2]. Our contribution here is aimed at the spatial monitoring of pH in a wound, relating this pH distribution to the underlying processes taking place. We also apply an external field which can affect the distribution of charges and charged objects in a wound in competition with the endogenous electric fields which are generated in all wound healing processes.

The pH of a wound is an aspect of the healing process that until recently has been given little attention. This although it is long known that biochemical processes and other physiological events are sensitive to the pH value of the environment [3,4]. During the last decades it has been shown that the wound healing process is associated with a spatiotemporal change in pH in the wounded area [5,6]. Such pH-profiles of wounds healing in a normal fashion are significantly different from those associated with healing processes in chronic wounds [6]. However a realistic description as well as a theoretical understanding of the spatiotemporal changes in pH is lacking in general. An initial attempt was given in [7] however it needs further refinement and theoretical development, which is addressed in this paper.

Recently Schreml et al. reported on a new technique of measuring the pH of a wound with high spatial resolution in the sub-mm range [5]. This opens for many new opportunities to study correlations and to understand the processes involved in the observed changes of the wound pH landscape. It should be emphasized that there is not one single value of pH which characterizes the whole wound, thus landscape is more appropriate. An improved understanding of this landscape could help further the development of diagnostic tools for assessing the progress of a healing wound as well as that of therapeutic methods in wound care where pH related aspects could be taken into consideration either directly or indirectly. This quest should be facilitated by correlating experimental results with a more basic theoretical model – of which we hope to build the base in this paper.

Below we develop a model of the pH in a wound not only to be able to make predictions but also to obtain insights about processes that are difficult or impossible to observe isolated or one at a time. We use the simplest model possible for the physiological events taking place and investigate the spatial changes in pH predicted by it. By comparing a set of scenarios to actual situations, we want to be able to judge the relative importance of in particular the three different factors; diffusion, wound electric field and boundary restrictions.

There are of course numerous factors influencing the healing process. Any such factor can delay or impede the healing process preventing the normal and optimal processes from occurring. Such factors can be extrinsic such as age of patient, chronic diseases, immunosuppression, nutritional status, smoking, alcohol abuse… Others are intrinsic ones like, ischemia or hypoxia, infections, foreign bodies, pressure, edema, necrotic tissue… [8,9]. To be able to come anywhere with a modeling approach this study is limited to investigating normally healing acute partial thickness wounds in cutaneous tissue. However a natural extension of this model would be to include some of the factors given above, once more detailed experimental data is available on the pH in real life situations.

In the next section we give some background to the major pH controlling factors in a wound as well as their influence on the wound healing process. Based on this we construct our model and discuss the outcomes of it in the subsequent sections.

## Wound healing and pH

pH has a long history from the first documented pH indicator in the 16th century (Leonard Thurneysser) to Søren Peder Lauritz Sørensens definition of the pH scale 400 years later. As a measure of the hydrogen ion activity of a solution it can for our purposes be approximated with the concentration *c* of protons in the solution: $c = [H^+]$, and pH = $-\log c$. Because of the logarithmic scale a rather precise regulation of pH can still correspond to large changes in concentration. Compared to other ions in biological systems, the concentration of $H^+$ varies over a large range, as given by the values for gastric acid pH=1, human skin 4-6, urine 5.7, cytosol 7.2, blood 7.4 and the Mitochondrial matrix 7.5. Virtually every biochemical reaction as well as cell and body functions are influenced by the pH value at hand [3,4]. On a macroscopic level cell shrinkage/swelling is related to pH [10] where tissue swelling is a normal state in the wound process. On a microscopic level the control of e.g. $H^+$ flows through pump-dependent changes in membrane voltages can be correlated with regeneration [11]. In this context we should stress that $H^+$ is not the only important factor. For instance hydrogen peroxide is a substance which can play a crucial role in wound healing; a reactive substance which can diffuse across cell membranes, within tissue and assembles at wound edges on time-scales of the order of minutes [12].

The effects of a change in pH can be severe since amino acids, donating or receiving a proton thus changing their surface charge, can change their conformation and hence their function and activity. This pH sensitivity, cells can also utilize for precise regulation of important reaction pathways. This is true both for intracellular and extracellular macromolecules which means that the microenvironment of the cells is influenced by and influences the local pH. We believe that pH-levels are even more significant when new tissue is formed in a wound cavity, for several reasons. First we do expect the cells in a wounded area to have an increased metabolism to provide the necessary energy and molecules needed in order to restore (pH) homeostasis; this makes up for large fluxes of protons. Secondly the increased metabolism results in an increased need for oxygen, while the blood supply to the tissue is decreased upon injury. This means there is less oxygen around to form water and hydrogen peroxide. This lowers the pH. Third, the production and remodelling of new tissue may also be associated with the production, and consumption of hydrogen ions. In this sense we would expect the wound to be a hydrogen sink.

Wound healing is often divided into three stages: inflammation, proliferation and maturation. In the initial stage of the inflammation phase the first minutes to hours after injury, the blood coagulates and forms a provisional plug. The wound would then have a pH very close to that of blood, about 7.4. But from that moment on, when wound healing progresses there is a change in pH. After about two weeks the wound is healed and the pH is restored to that of skin. This wound healing scenario has to do with the proliferation and migration of different cell types which is also related to (local) pH. It is found that the optimal pH for the proliferation of two cell types in wound healing, keratinocytes and fibroblasts is between pH 7.2 and 8.3 [13,14]. Both cell types rely on the function of the enzymes matrix metalloproteinases. These are zinc dependent protein degrading enzymes that play a vital role for migration as migrating cells must degrade the extracellular matrix in front of them in order to be able to move forward. These enzymes have pH optima in the vicinity of 8 *in vivo*. In a recent study, skin like cells, were found to orient and migrate along an extracellular pH gradient [15,16].

Of particular interest for us is mapping the spatial distribution of pH – especially pH -gradients. In the wound situation there will be new ways for protons to enter or leave the wound fluid compared to the situation before. The wound edge and bottom provides for the meeting between a buffering region and the wound fluid. We would anticipate large possible pH gradients. Of course even in a normal situation there is a pH gradient spanning the whole of the epidermis setting the length-scale of a normal pH-gradient in the skin system [15-17]. The skin has a pH of about four to six, depending on where the pH is measured, individual differences, age, health status etc. [15] Skin being acidic is important for its barrier function as many pathogenic bacteria have a pH optimum that is more alkaline [6,8,12]. Acidic micro-domains (average pH 6.0) of the order of µm in size can be found in the extracellular matrix of the stratum corneum However in mid-stratum corneum the intracellular space approaches a pH of 7.0. As we traverse epidermis the pH quickly increases. The concentration difference of protons over this very thin strata of tissue is as high as 100-1000 fold [15,16]. Neutrality is reached about 10 µm below the skin surface, and at the stratum basale the pH is about 7.5-8 [15,18]. Blood vessels are frequent in dermis and still the pH in dermis is lower than that of blood. This would suggest that the maintenance of dermis involves a slight net consumption of hydrogen ions.

We now turn to another important ingredient in our modelling since the appearance of a wound is associated with strong local electric fields which can influence the proton balance and hence pH.

## Wounds, pH and electric fields

When considering the movements of protons, being charged particles, in and near a wound it is important to have an insight in the electric fields acting. As a potential difference ranging from 10 to 60mV [19-21] is set up between *stratum corneum* and *stratum basale*, known as the trans-epidermal potential, this gives rise to an electric field of the order of 100V/m for a typical thickness of epidermis of the order of a couple of hundreds of micro-meters. Through the combined efforts of ion transporters and pumps, there is a net transport of chloride ions towards the surface of the skin and sodium ions in the opposite direction: towards *stratum basale* and dermis. The interface between epidermis and dermis is therefore rendered a positive potential relative the surface of the skin and the electric field points towards the skin surface.

Epidermis is ruptured when a wound is formed, then short-circuiting the trans-epidermal potential [22,23]. The potential collapses to zero in the centre of the wound as the mechanisms setting up the potential are no longer functioning and this affects the trans-epithelial potential up to a millimetre from the wound edge [23]. The electric field between *stratum basale* and *stratum corneum* is thus reoriented close to the wound edge now pointing towards the centre of the wound [24,25]. The electric field dies quickly off when we move from the wound edge into the wound [23] and this length-scale will be an important one for our pH modelling as we will see below.

The electric field described above exerts a force on ions present in the wound area, such as protons. Positive ions migrate towards the wound. The new orientation of the field lines as a consequence of skin rupture is believed to be important by giving a sense of orientation to the wound healing phase. The unperturbed field in skin pointing towards the surface is also consistent with having a lower pH at the skin surface compared to *stratum basale*.

There are a set of pH-altering reactions that take place at the surface of a wound if it is exposed to the surrounding atmosphere. These include loss of carbon dioxide, drying of the surface and oxidation reactions [26]. The lost and damaged tissue needs to be replaced. This is achieved by the cellular production of extracellular matrix components, such as the protein collagen. These have a buffering effect and stabilizes pH, and hence may significantly lower the diffusion coefficient of the protons as well as the amount of free protons hereby lowering the proton activity. A low pH is known to stimulate new blood vessel formation. Thus the buffering capacity of blood vessels that exists or are formed during the wound healing process also contribute to the buffering of the wound. Hence a wound acts as either a sink or a source depending on the pH it has and the considered endogenous electric fields will also influence the detailed physics taking place.

In these first three sections we have looked into the reasons as to why pH is important to wound healing and hereby motivated its relevance in wound care. We have also seen that yet there is no quantitative data on the spatial change in the pH of a wound but that the techniques to obtain such data recently have been developed so that we can expect to have such experimental information within the near future. We have also looked at some possible sources and sinks of hydrogen ions that are associated with wound healing and some properties of proton mobility in aqueous solutions with an electric field present. This knowledge is put to the test in the next section where we describe how a model to calculate the pH in a wound can be developed including the local electric field acting in the wound area.

## Description of model

The spatiotemporal change in the pH of the wound is now formulated within a model containing the main features described above. The model is supposed to be compared to experimental results in order to validate it and develop it further; data which is not at hand at present. The model as such will however define the relevant length, time and energy scales in the problem having implications for what variables to change and how much, in order to achieve a particular response. It also yields insight into what contributions we can expect from the field part as well as from the source part.

There are many feasible ways to go about modeling wound healing. Numerous aspects of wound healing have been modeled such as transport or diffusion of different species such as growth factors, wound closure velocity and the influence of geometry [27,28]. There is only one theoretical study dealing with the change in pH of the kind we address in this paper [7]. However whereas we focus on spatial scales [7] focusses on the time-scale.

The parameters we will use are based on physiological events and our formulation is in terms of conserved quantities by starting from the continuity equation for the proton concentration. The wound bed is regarded as a volume surrounded by tissue of a constant pH. The reason for this is that the viable tissue surrounding the wound has a good buffering capacity because of the undisturbed blood flow perfusing it. The change in the number of protons at each point in the wound is determined by the net flux of ions to or from that point and the generation or loss of ions at that position. As the skin is ruptured, the proton movement to the damaged area (ion flux) is dramatically changed as compared to the situation prior to the injury. Wound fluid perfuses the domain and new routes (mainly in the wound) are created for ions to move so as to minimize their potential energy owing to the potential difference between the *stratum corneum* and *stratum basale*, as described in the previous section. The major part of the endogenous electric field now points in an orthogonal direction compared to the situation prior to wounding; i.e. along the wound bed pointing towards the wound center.

Integrating the continuity equation over a reasonable sized domain, the change of protons within it is represented by the divergence of the net flux of ions over the boundaries *j(x,t)* as described in Eq.(1) below. The domain in our system is the wound bed and for the hydrogen concentration *c(x, t)* we have:

$$\frac{\partial c}{\partial t} + \nabla \cdot \boldsymbol{j} = \lambda(\boldsymbol{x}, t) \quad , \tag{1}$$

where $\lambda(\boldsymbol{x}, t)$ is a source/sink accounting for if hydrogen ions are liberated or consumed in the space point *x* at time *t*. In terms of concentration gradients (for a diffusion constant *D*) and an applied electric field *E(x,t)* the flux can be related to the concentration through:

$$\boldsymbol{j}(\boldsymbol{x}, t) = -D\big(\nabla c(\mathrm{x}, t) - \beta q c(\boldsymbol{x}, t) \boldsymbol{E}(\boldsymbol{x}, t)\big) , \tag{2}$$

in which q is the charge of the carriers, we have defined $\beta \equiv 1/k_B T$ where *T* is the absolute temperature, and used the Einstein relation to express ion mobility in terms of the diffusion constant.

To continue we introduce some simplifying assumptions, apart from the major one of only considering protons as indicated above; rendering an equation which captures the main spatio-temporal change of the pH we are looking for. We assume the wound is rotationally symmetric and shallow enough that we only consider what happens at the wound edge and in the now two-dimensional wound bed. Hence the general function $F(\boldsymbol{x}, t) \equiv F(\rho, t)$, i.e. the only spatial dependence in the problem is the radial coordinate $\rho$ from the center of the wound.

Based on the discussion in the previous paragraph we write the electric field as $\boldsymbol{E}(\boldsymbol{x}, t) = E_o g(\rho)(-\hat{\rho})$ where $E_o$ is a measure of the field strength, *g* describes the spatial shape of the electric field in the wound and the last factor expresses that the field points in the negative radial direction, i.e into the wound. We assume the proton time-scale for diffusion is fast compared to the time-scale of upholding the trans-epithelial potential why we neglect the influence of protons and other ions on the electric field itself. Thus we neglect the time-dependence in $\boldsymbol{E}(\boldsymbol{x}, t)$ in what follows.

Re-scaling Eqs. (1-2) in terms of a length-scale $L=k_B T/qE_o$ and a time-scale $T=L^2/D$ we get the following equation:

$$\frac{\partial \bar{c}(\bar{\rho}, \bar{t})}{\partial \bar{t}} = 4\bar{\lambda}(\bar{\rho}, \bar{t}) + \left[\frac{1}{\bar{\rho}} \frac{\partial}{\partial \bar{\rho}} \bar{\rho}\right] \left(\frac{\partial \bar{c}(\bar{\rho}, \bar{t})}{\partial \bar{\rho}} + g\bar{c}(\bar{\rho}, \bar{t})\right). \tag{3}$$

The concentration c is normalized to a reference concentration $c_o$ which we define as the one of the surrounding tissue as obtained from its pH-value $pH_{surrounding}$ = - log $c_o$ . The source $\lambda$ is normalized to a source of strength $4c_o/T$. The factor of four is for later convenience.

Let us pause briefly and investigate these scales governing Eq.(3). At room and body temperature $k_BT/q$ is around 25 mV for a unit charge like a proton. With $E_o$ of the order of 100V/m this yields a length-scale L of the order of millimeter or less. For a distance L, the diffusion flux driven by the thermal energy, is equal to the flux caused by the electric field. When the wound is larger than L, the shape and extension of the electric field becomes important. If it is smaller than L, then diffusion dominates. L is thus one possible parameter do decide when a wound is "large or small" the other one being the physical wound size b as we will denote it in what follows.

The time-scale in our problem, $T=L^2/D$, corresponds to the time it takes for a concentration change to diffuse over the distance L. For a typical D for protons in water $10^{-8}$ m$^2$/s, this gives a time scale of the order of 10 seconds. This is short compared to the time-scales of over-all wound healing (days) as well as cell division times (many hours). We also notice that for a typical cell movement velocity of 10 μm/h in would healing, this time-scale is so long that very little cell-movement has taken place during it. When it comes to the characteristic source term $4c_o/T$ we can do the following estimate. For a pH of 7 in surrounding tissue and T of the order of 10 s this corresponds to a characteristic source strength of the order of 10 nM/s which is our measure if a chemical perturbation is small or large in the system.

Since we have a situation where time-scale is set by processes which are very much faster compared to the typical time-scale of the wound healing process we are not interested in transients or fluctuations in the density rather its stationary value. This means that effectively we have a quasi-static problem and can drop the time-dependence in the concentration. We can then parameterize our boundary b as b(t) where t is the wound healing time-scale and at each instance in time move the boundary without immediately affecting the concentration or changing the equilibrium situation in the surroundings. The stationary part of Eq.(3) is:

$$\frac{\partial \bar{c}(\bar{\rho})}{\partial \bar{\rho}} + g(\bar{\rho})\bar{c}(\bar{\rho}) + 2\bar{\rho}\bar{\lambda}(\bar{\rho}) = 0 \ . \tag{4}$$

For convenience we will from now on use a <u>constant</u> source density given the little information we have at hand for choosing anything else at present, mainly discussing in terms of a source or sink. Notice that the field strength and temperature effects are completely absorbed in the length-scale L of the problem, while the field shape is present through the g-factor. We furthermore notice that it is the source density near the perimeter which is most important being multiplied with the factor $\bar{\rho}$ in Eq.(4). For a sink ($\bar{\lambda}$ negative) we see from Eq.(4) that the normalized concentration has to increase when approaching the boundary. For a source it decreases when approaching the boundary (where $\bar{c} = 1$).

The general solution of Eq.(4) can be written in the following form:

$$\bar{c}(\bar{\rho}) = G(\bar{\rho}, \bar{b}) + 2\bar{\lambda} \int_{\bar{\rho}}^{\bar{b}} dp\, p\, G(\bar{\rho}, p), \qquad (5)$$

defining the electric field profile function

$$G(a, b) = e^{\int_a^b dp\, g(p)}, \qquad (6)$$

as an average of the field profile function *g(ρ)* over parts of the wound bed. Notice that the source contribution to the concentration is a first moment of the source-free concentration.

It is a straightforward matter to calculate the concentration profile for different electric field profiles. However without a real experimental guidance it is easy to generate results *ad infinitum* with little relevance to the reader. Therefore we look at four simplifying situations to gain an increased insight in what types of pH profiles which can be generated and what is the cause behind them.

**A.** If we have *no source* present $\bar{c}_{no\, source}(\bar{\rho}) = G(\bar{\rho}, \bar{b})$, where $\bar{b}$ denotes the boundary and is typically very large for wounds exceeding a few millimeters. It is a simple matter to show how the field profile function *g* then determines the pH difference in the wound:

$$pH_{no\, source}(\bar{b}) - pH_{no\, source}(\bar{\rho}) = \log \bar{c}_{no\, source}(\bar{\rho}) = \int_{\bar{\rho}}^{\bar{b}} dp\, g(p)/\ln 10. \qquad (7)$$

Since for most relevant situations *g* goes between *0* and *1* between $\bar{b} - 1$ and $\bar{b}$, where we remind the reader the lengths are measured in units of the characteristic electric length scale *L=k_BT/qE_o* , the left hand side of Eq.(7) indicates that we can achieve a pH-difference of no more than around a unit. This is so because $\bar{b}$ drops out of the problem and we see the inside is slightly more acidic than the wound boundary. This is the large wound limit. For a small wound ($\bar{b} < 1$) where the field is roughly constant throughout the pH-difference is proportional to $\bar{b} - \bar{\rho}$ again implying a small change in pH, now also at the origin. As the wound heals the distance between the boundary $\bar{b}$ and the middle decreases continuously.

**B.** In light of the result in **A.** in order to have a substantial change in pH we would have to complement the endogenous electric field with an exogenous one so the total field exists all over the wound extension. Thus if we assume we can create a constant external field all over the wound bed, Eq.(7) tells us that we can achieve a significant change of pH. Its largest value is at the origin and is proportional to $\bar{b}$ where $\bar{b}$ in the new situation can be of substantial size and we easily get changes in pH by one unit or more.

On the other hand if we want to introduce an external electric field for other reasons and not to change the pH itself dramatically, and there are as outlined in the introduction several positive benefits associated with this, one could choose a field profile g such that the integral in Eq.(7) is as small as possible. One such electrical dressing is already available in the market today where g alternates in sign over dimensions of the order of millimeters; that is several units in terms of our basic length scale *L*. It has been shown that it makes re-epithelialization and keratinocyte migration going faster [29] than with standard treatments. In this way one can achieve a situation with no change in over-all pH however on the scale of millimeters change its value creating local pH gradients which are known to influence cell response [15,16]. As mentioned earlier Keratinocytes e.g. exhibit a migratory phenotype at high pH, proliferate at a range of pH values and differentiate at low pH. [14]

C. If we have no electric field present at all, $G \equiv 1$, we get:

$$\bar{c}_{no\ field}(\bar{\rho}) = 1 + \bar{\lambda}[\bar{b}^2 - \bar{\rho}^2], \tag{8a}$$

showing a quadratic dependence with respect to radius. In this case our length-scale *L* has no meaning instead it is the wound radius *b* which is the characteristic length-scale to be used. We therefore write Eq.(8a) in the following way:

$$\bar{c}_{no\ field}(\bar{\rho}) = 1 + \frac{A(b) - A(\rho)}{A_\lambda}, \tag{8b}$$

defining $A(\rho) \equiv \pi\rho^2$, the area for a circle of radius ρ and $\lambda A_\lambda \equiv 4\pi D c_o$. Thus the concentration at a certain radius depends on the wound area beyond that point. Furthermore the factor $4\pi D c_o$ is closely related to the diffusion flux to an absorber and is a constant for our problem. For small λ the pH-profile deviates very little form the pH in the surrounding tissue as it should. At the center of the wound the concentration is $\bar{c}_{no\ field}(\hat{\rho} = 0) = 1 + A(b)/A_\lambda = 1 + \lambda/\lambda_o$, where $\lambda_o = 4Dc_o/b^2$ is the characteristic source strength for a system of size b.

D. It is not unlikely that the source-term itself depends on the concentration; i.e. the buffering of protons depends on the proton concentration itself. We therefore make the plausible assumption $\bar{\lambda} = \bar{\lambda}_{av}\bar{c}(\bar{\rho})$ for illustrational purposes. Eq.(4) can then be written:

$$\frac{\partial ln\bar{c}(\bar{\rho})}{\partial\bar{\rho}} + g(\bar{\rho}) + 2\bar{\rho}\bar{\lambda}_{av} = 0. \tag{9}$$

For a constant electric field near the wound edge we then find:

$$(pH(\bar{b}) - pH(\bar{\rho}))/loge = \gamma_{\bar{b}-\bar{\rho}} + \bar{\lambda}_{av}(\bar{b}^2 - \bar{\rho}^2) \ . \tag{10}$$

$\gamma_{\bar{b}-\bar{\rho}}$ is an average value of the g-factor thus of the order 1. It is constant throughout the wound until we reach the wound edge where it drops to zero as $(\bar{b} - \bar{\rho})$. Eq.(10) shows again a quadratic dependence for the source part when approaching the middle of the wound while the electric part is constant for most of the wound space in accordance with the other findings above. We can of course manipulate the g-factor with external means to change this.

## Concluding remarks

Our major concern in this paper has been to gain an increased understanding for the spatial pH-distribution in a wound with only internal electric fields or applying external ones. Experimental reports on this are scarce, however recently Schreml. et. al. [5] published a study with sub-millimetre resolution and more detailed results are therefore anticipated in the near future.

We have tried to develop a model to find relevant spatial scales and to yield an insight in the behaviour of pH in a confined environment. Natural electric fields of the order of 100V/m have been invoked. They are important since they affect the generation of e.g. hydrogen peroxide which when increased in concentration induces cell migration by coupling to the cell cytoskeleton through appropriate membrane proteins.

Given the time-scales in our problem it was turned into a stationary one and we mimicked the wound healing process by changing the boundary slowly with time as in real wound healing.

Most of our major observations are most easily expressed by integrating Eq.(4) between a certain position in the wound and the boundary to generate an integral equation for the concentration. This gives us the general result:

$$\bar{c}(\bar{\rho}) = 1 + \bar{\lambda}(\bar{b}^2 - \bar{\rho}^2) + \int_{\bar{\rho}}^{\bar{b}} dp \ g(p) \ \bar{c}(p) \ , \tag{11}$$

showing in a more explicit way the boundary condition (first term), the sink/source second term and the third term the electric field contributions. The latter one has a very weak spatial dependence in the endogenous situation where g is non-zero only in a very small interval next to the boundary. Therefore the term is basically a constant, of order unity, up to $\bar{\rho} = \bar{b} - 1$ and then drops to zero at $\bar{\rho} = \bar{b}$. In this sense the major factor influencing the pH is the source density which is geometrically boosted by a factor $\bar{b}^2$ as detailed in example **C** in the previous section. This implies that a major possible change of pH in our wound model is a number of the order of $ln\bar{b}$. This is one of the major results of this work - the pH in a wound situation depends directly on the area of the wound itself. This might in itself be a problem since it is governed by the geometry at hand which usually cannot be influenced to any significant degree and one has to resort to changing the source density if possible. However there is another possibility of trying to divide a wound in smaller units like introducing a skin graft. A large wound divided in smaller units make every unit to be a small wound where pH cannot change significantly from the surrounding. Furthermore manipulating the field profile with an external electric field one can change this situation to suit the needs which have to be met to facilitate a good wound healing process. This means that shaping the g-factor by external means. For instance so called *Electroceuticals* can be designed to have electric field properties which could enhance or influence the pH one think is important in a special situation.

In our work we did not look into the presence of bacteria which is also related to pH. As previously mentioned, the low pH of skin is believed to be important for its barrier function. This as growth of most human pathogenic bacteria is inhibited at pH values below six. There is also a feed-back loop here: bacterial colonization has been shown to be correlated with an increase in skin pH. Hence, the pH of a wound is important to the basic conditions for a colonization of pathogenic bacteria and once in the wound, they may alter the pH in the wound.

We hope that our modelling is a first small step in facilitating a deeper understanding of the role of pH in wound healing.

## Acknowledgment

P. Apell acknowledges enlightening discussions with Bo Ekman and the generous atmosphere provided by Ludvig Lizana at the Ice Lab, Umeå University.